\documentclass[10pt]{article}
\usepackage{color}
\usepackage{units}
\usepackage{upgreek}
\usepackage{cite}
\usepackage[parfill]{parskip}
\usepackage[onehalfspacing]{setspace}
\usepackage{graphicx, amsmath, amssymb, amstext, enumerate, subfigure, listings, longtable, lastpage, fancyhdr, fancybox, hyperref}

\begin{document}

\begin{doublespace}
	
\textsc{{\LARGE  A near-field study on the transition from localized to propagating plasmons on 2D nano-wedges}}

\end{doublespace}

\textbf{Thorsten Weber,$^{1,2}$ Thomas Kiel,$^{3}$ Stephan Irsen,$^{2}$ Kurt Busch$^{3,4}$ and Stefan Linden$^{1,\ast}$}

\textsl{
$^{1}$Physikalisches Institut, Rheinische Friedrich-Wilhelms Universit\"at, 53115 Bonn, Germany\\
$^{2}$Electron Microscopy and Analytics, Center of Advanced European Studies and Research, 53175 Bonn, Germany\\
$^{3}$Institut f\"ur Physik, Humboldt-Universit\"at zu Berlin, 12489 Berlin, Germany\\
$^{4}$Max-Born-Institut, 12489 Berlin, Germany\\
$^{*}$linden@physik.uni-bonn.de} 

\textbf{Abstract:}\\
In this manuscript we report on a near field study of two-dimensional plasmonic gold nano-wedges using electron energy loss spectroscopy in combination with scanning transmission electron microscopy, as well as discontinuous Galerkin time-domain computations. With increasing nano-wedge size, we observe a transition from localized surface plasmons on small nano-wedges to non-resonant propagating surface plasmon polaritons on large nano-wedges. Furthermore we demonstrate that nano-wedges with a groove cut can support localized as well as propagating plasmons in the same energy range.



\section{Introduction}
Focusing light down to nanometric volumes is of great interest for various applications. For instance, nanoscale spectroscopy and imaging techniques such as tip-enhanced Raman scattering, apertureless near-field optical microscopy (A-NSOM), and ultrafast photoemission of electrons make use of highly confined electromagnetic fields \cite{Hartschuh2003, Keilmann2004, Neacsu2010, Kruger2011, Herink2012}. 
Many of these techniques rely on the excitation of plasmonic modes in suitable metallic nano-structures, where the collective oscillations of the conduction band electrons in the metal lead to strongly enhanced electromagnetic near-fields\cite{Maier2007}. 

In metallic nano-structures that are small compared to the relevant free-space wavelength, e.g. nano-antennas \cite{Muehlschlegel2005}, plasmonic oligomers \cite{Hentschel2010}, or split-ring resonators \cite{Boudarham2010}, the boundary conditions give rise to standing wave patterns of the charge carrier oscillations. The corresponding resonant modes are the so-called localized surface plasmons (LSPs), which typically exhibit hot-spots of the electromagnetic field in the vicinity of the surface. The  spectral and spatial properties of the LSPs can be controlled by designing the shape, size, or surrounding media of the nano-structure \cite{Maier2007}. 

Extended metal nano-structures support propagating surface plasmon polaritons (SPPs), which are formed by charge carrier density waves moving along the surface. On tapered nano-structures such as nano-wedges, the SPP dispersion relation varies along the propagation direction. This can be utilized to compress SPPs and to achieve strong local-field enhancements. For instance, a SPP propagating along a nano-wedge towards the apex is gradually slowed down as it reaches the apex region \cite{Nerkararyan1997, Stockman2004}. For most experimentally accessible wedge parameters, the SPP is however not brought to a complete stop but rather is reflected at the apex\cite{Schroder2015, Yalunin2016, Guo2016}.

Precise knowledge of the spatio-spectral distribution of the plasmonic near-field is of utmost importance for the applications mentioned above. A very powerful experimental method that allows to map localized as well as propagating plasmon modes is electron energy loss spectroscopy (EELS) in combination with scanning transmission electron microscopy (STEM)  \cite{Nelayah2007,Bosman2007,Boudarham2010,Cube2011,Rossouw2011,Huth2013,Schoen2015,Schroder2015,Walther2016,Cube2013}.  In STEM-EELS, a highly focused electron beam scans over the sample. Passing nearby or through the structure, the electrons can excite plasmon modes. As a result, the fast electrons lose kinetic energy by interacting with the self-induced electric field in the vicinity of the nano-structure. The probability for this process, the so called electron energy loss probability (EELP), is related to the plasmonic local density of states (LDOS) and can thereby be used to characterize the plasmonic nano-structure \cite{GarciadeAbajo2008,GarciaDeAbajo2010,Hohenester2009}.

In this work, we report on a near field study on two-dimensional plasmonic gold nano-wedges using STEM-EELS. With increasing nano-wedge size, we observe a transition from LSPs on resonant nano-wedges to non-resonant propagating SPPs on large nano-wedges. Furthermore we demonstrate that nano-wedges with a groove cut can support both, localized and propagating plasmon modes.
The experimental data is compared to numerical computations based on the Discontinuous Galerkin Time-Domain (DGTD) method.

\section{Methods and instrumentation}
All samples have a wedge-like form with an opening angle of $14^\circ$ and are fabricated by standard electron beam lithography on $\unit[30]{nm}$ thin silicon nitride membranes \cite{Cube2011}. The structures themselves consist of $\unit[30]{nm}$ thin thermally evaporated gold films. 

For the STEM-EELS experiments we use a Zeiss Libra200 MC Cs-STEM CRISP (Corrected Illumination Scanning Probe) operated at $\unit[200]{kV}$. The instrument is equipped with a monochromated Schottky-type field-emission cathode and a double hexapole-design corrector for spherical aberrations of the illumination system (Cs-corrector).  An in-column omega-type energy filter, fully corrected for second-order aberrations, is integrated into the microscope. The spectra are recorded using a Gatan UltraScan 2k x 2k CCD camera with an acquisition time for each spectrum of $\unit[5]{ms}$ and a dispersion of the spectrometer of  $\unit[0.016]{eV}$ per channel. The energy resolution, which is defined by the FWHM of the spectrum's zero-loss peak measured through the silicon nitride membrane, is $\unit[0.1]{eV}$ in the center of the scan area. In the present STEM-EELS experiments, the electron beam raster scans over the sample with a step width ranging between $\unit[5]{nm}$ and $\unit[20]{nm}$, depending on the size of the investigated area. 
All spectra are recorded for normal incidence of the electron beam. 
For data postprocessing, each spectrum is normalized to its total number of electron counts. Subsequently, the first moment of the zero-loss peak is centered to $\unit[0]{eV}$ and a background spectrum is subtracted. In the presented EEL maps, the signal is normalized to the maximum value found in the relevant area for the given electron loss energy.


\begin{figure}
	\centering
	\includegraphics[width=\linewidth]{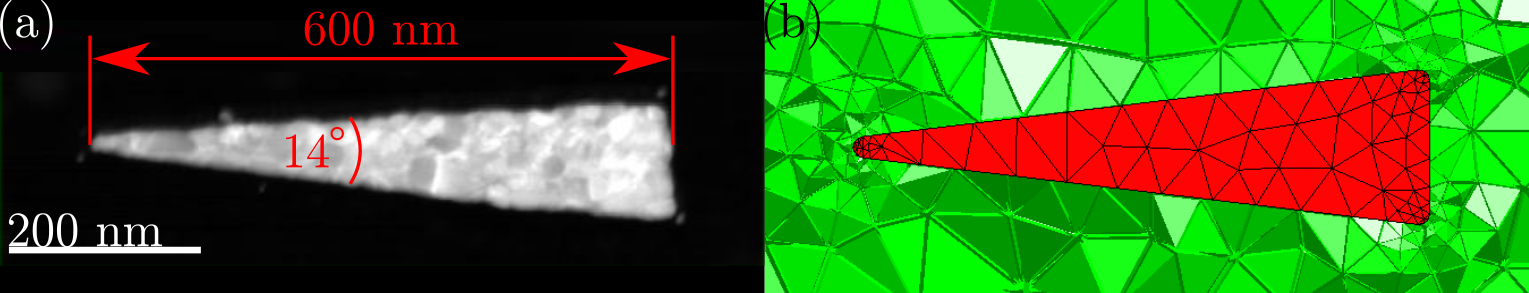}
	\caption{(a) Dark field micrograph of an investigated nano-wedge. (b) Detail of the mesh used for computing the data shown in figure~\ref{DGTDsmall}. The mesh's dimensions are read out from (a).}
	\label{mesh_short}
\end{figure}

To support our experimental findings we have performed numerical computations for selected structures using the DGTD method\cite{Busch2011,Matyssek2011}. 
The geometrical parameters for the nano-wedge are taken from the corresponding HAADF micrographs.
The nano-wedges are embedded into vacuum and have a dielectric permittivity, which is approximated using a Drude-Lorentz-model for gold similar to Ref.~\cite{Schroder2015}.
The swift electron's speed is set to $v = \unit[0.77]{c}$.
We obtain the induced polarization of a relativistic moving electron passing the gold nano-wedges and determine the back-action of the induced field on this electron \cite{Ritchie1957,GarciaDeAbajo2010,Matyssek2011}. 
The expansion of the electric and the magnetic field into Lagrange polynomials of third order is carried out for each mesh element of the structure. The mesh's geometry is determined by inspection of the micrograph of the corresponding structure and the mesh's elements have side lengths down to $\unit[5]{nm}$ (see figure~\ref{mesh_short}).

\section{Results}
\begin{figure}[htb]
\centering\includegraphics[scale = 0.83333]{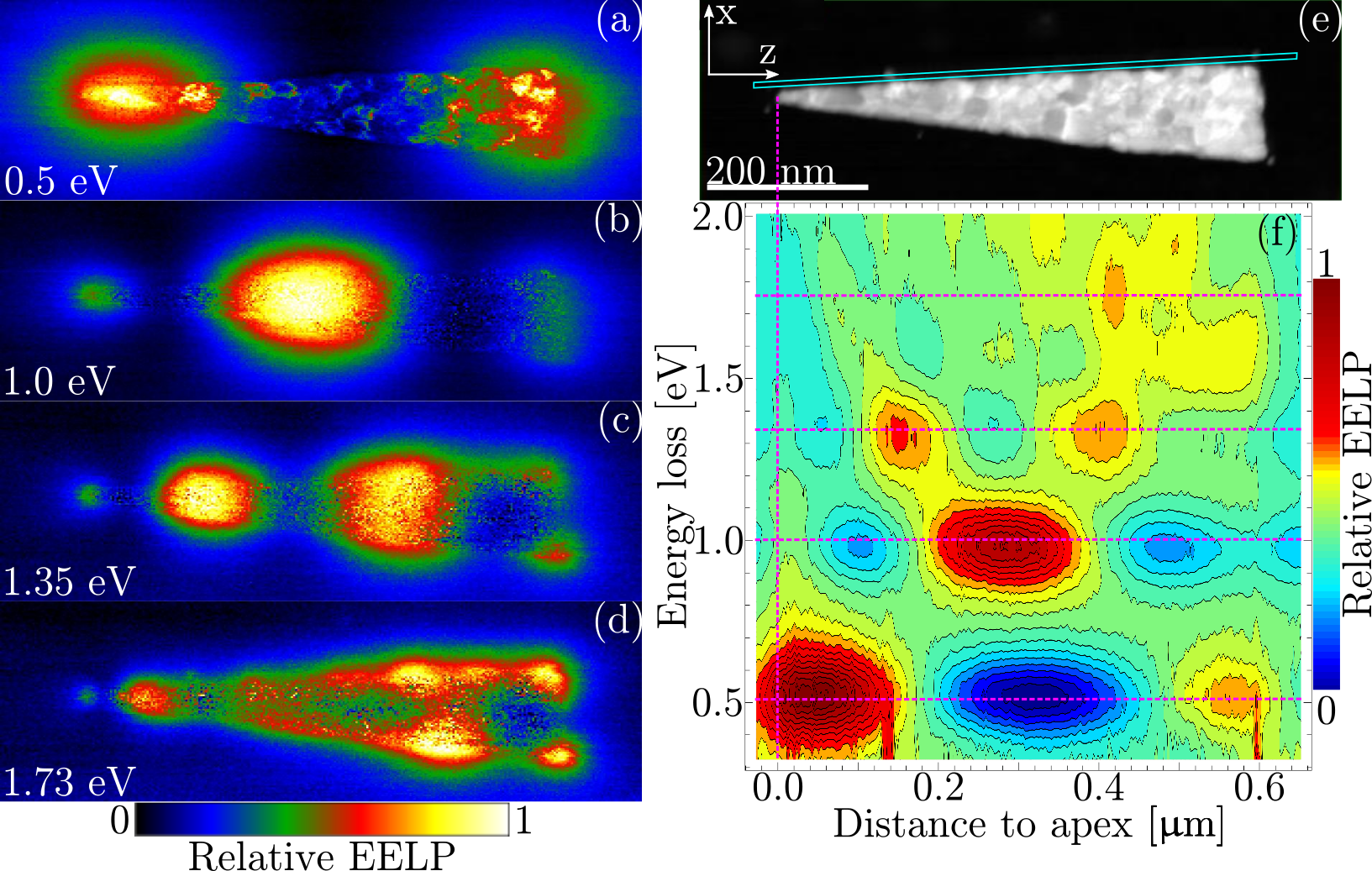}\caption{(a) -- (d): EEL maps of a $\unit[600]{nm}$ long gold nano-wedge for electron loss energies of $\unit[0.5]{eV}$, $\unit[1.0]{eV}$, $\unit[1.35]{eV}, $and $\unit[1.73]{eV}$. (e): HAADF micrograph of the investigated structure. The blue box indicates the area where the EELP data in (f) has been recorded. Red doted lines in (f) indicate the apex of the nano-wedge and the energies of the shown EEL maps.}
\label{short}
\end{figure}

As an example for a small resonant plasmonic nano-structure we first investigate a gold nano-wedge with a length of $\unit[600]{nm}$ (see electron micrograph in figure~\ref{short}).
As expected, the nano-wedge exhibits a number of distinct LSP-modes with resonance energies of $\unit[0.5]{eV}$, $\unit[1.0]{eV}$, $\unit[1.35]{eV}, $and $\unit[1.73]{eV}$, respectively. 
LSP-modes with larger resonance energies can not be clearly resolved because of the onset of the interband transitions in gold.  
The EEL maps of the observed LSP-modes are displayed on panels (a)-(d) of figure~\ref{short}. 
The fundamental LSP-mode ($\unit[0.5]{eV}$) has two pronounced EELP maxima near the ends of the nano-wedge.  This clearly indicates that this mode is of dipolar character.
The higher-order LSP-modes exhibit an increasing number of EELP maxima along the nano-wedge axis ($z$-axis). 
In a simple standing wave picture, the maxima correspond to the locations of nodes of the the charge carrier oscillations. 
Note that these LSP-modes could be excited in an optical far-field experiment with light polarized along the $z$-axis.

Figure~\ref{short}(f) displays the EELP recorded along the edge of the nano-wedge (see blue box in micrograph) for different loss energies on a false color scale. Here, the EELP signal was integrated over 3 pixels perpendicular to the edge for each $z$-position along the nano-wedge. 
In this representation, we can clearly identify the different discrete LSP-modes.

\begin{figure}[hbt]
\centering\includegraphics[scale = 0.83333]{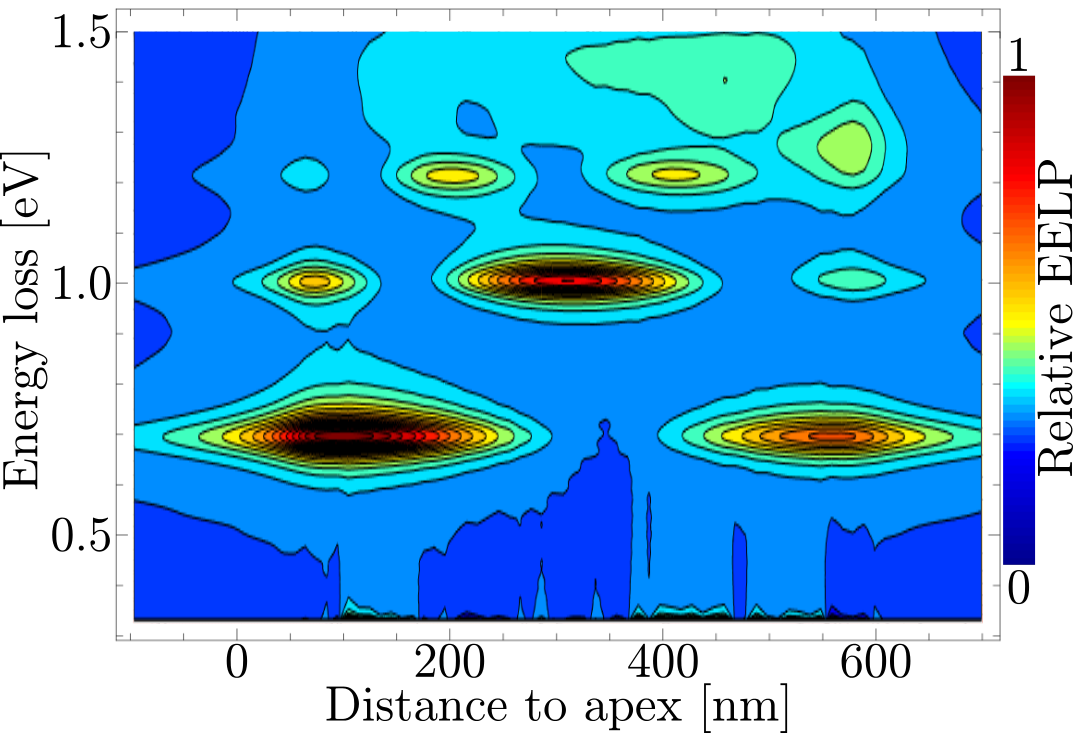}\caption{Relative EELP near a $\unit[600]{nm}$ long nano-wedge color coded as a function of distance from the apex and energy loss obtained from DGTD computations. The EELP is recorded along the upper wedge of the nano-wedge.
}\label{DGTDsmall}
\end{figure}

Our experimental findings are reproduced by the numerical computations. Figure~\ref{DGTDsmall} displays the computed EELP along a line parallel to the edge of a $\unit[600]{nm}$-long nano-wedge. The thickness of the gold film as well as the opening angle of the nano-wedge have been chosen as in the experiment. Like in the experimental data, we observe a set of discrete modes, which can be assigned to the localized surface plasmons of the nano-wedge. The offset in energy of the modes between experiment and numerical computations is most likely caused by deviations of the fabricated structure from the geometry assumed in the computations and a shift in energy caused by the influence of the underlying substrate, which was not taken into account in the computations.

A qualitatively different behavior is expected for metal wedges whose lateral dimensions are large compared to the plasmon decay length. 
In this case, a discrete standing wave pattern of the charge carrier oscillations can not form since multiple reflections from the nano-wedges's ends are suppressed by absorption. Hence, long nano-wedges are expected to support a continuum of propagating SPPs instead of discrete LSPs.

\begin{figure}[hbt]
\centering\includegraphics[scale = 0.83333]{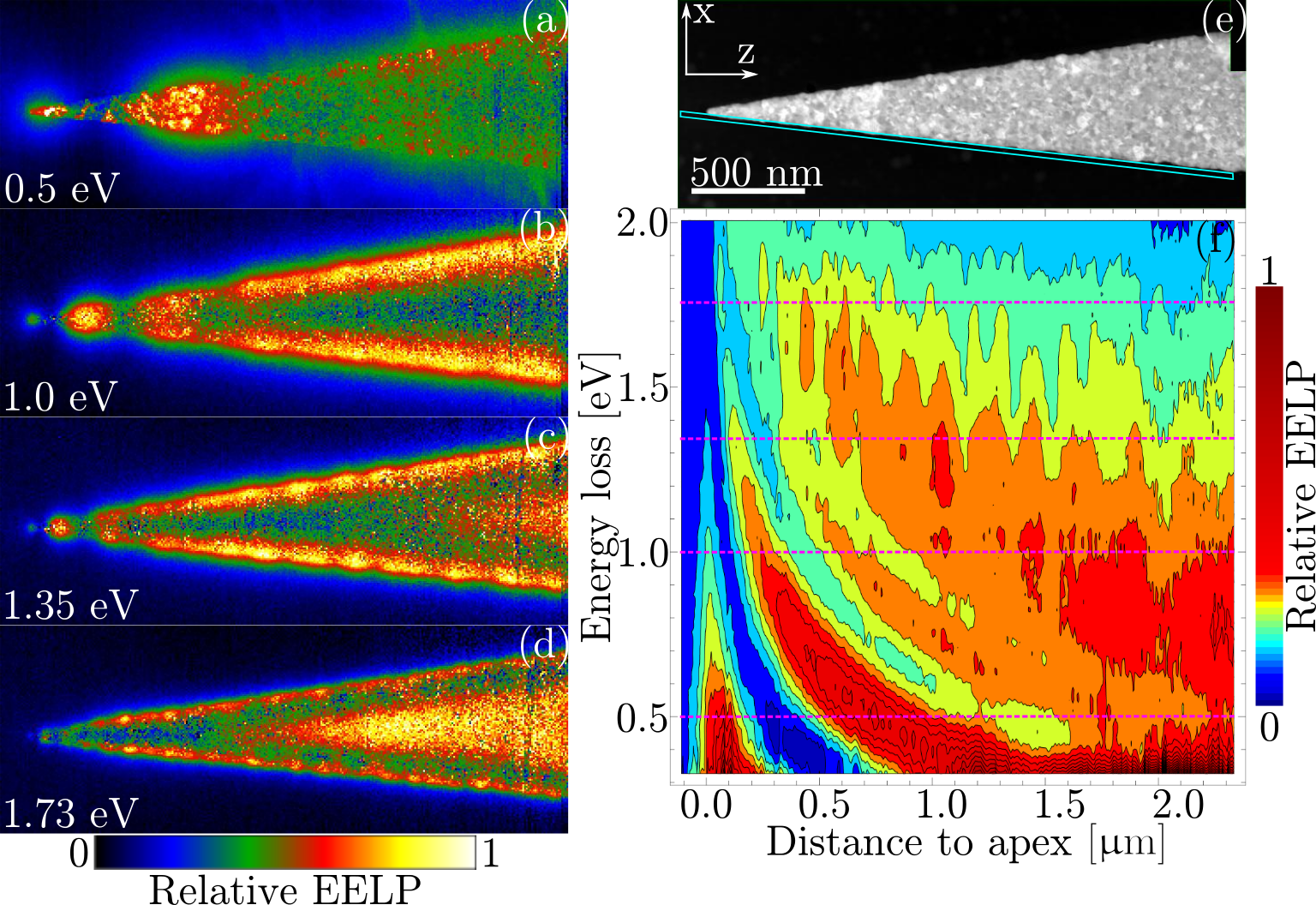}\caption{(a) -- (d): EEL maps of a $\unit[15]{\upmu m}$ long gold nano-wedge for electron loss energies of $\unit[0.5]{eV}$, $\unit[1.0]{eV}$, $\unit[1.35]{eV}, $and $\unit[1.73]{eV}$. (e): HAADF micrograph of the front section of the investigated structure. The blue box indicates the area where the EELP data in (f) has been recorded. Red doted lines in (f) indicate the apex of the nano-wedge and the energies of the shown EEL maps.
}\label{long}
\end{figure}

As an example for a non-resonant structure, we consider a $\unit[15]{\upmu m}$ long gold nano-wedge. 
The panels (a)-(d) of figure~\ref{long} display EEL maps of the first $\unit[2]{\upmu m}$ long part of this nano-wedge for the same electron loss energies as displayed in figure~\ref{short}. 
In contrast to the short nano-wedge, the long nano-wedge shows a non-resonant behavior. The strong signals at the edge and in the interior of the nano-wedge correspond to the SPP edge modes and the film modes, respectively, as studied by Schmidt et al.\cite{Schmidt2015} on 2D silver tapers. 
In the following, we will concentrate on the SPP edge modes.
For this purpose, we consider the EELP recorded along the lower edge of the nano-wedge (see blue box in figure~\ref{long}(e)). This data is displayed for different loss energies on a false color scale in figure~\ref{long}(f).
The EELP exhibits at all loss energies a maximum near the apex of the nano-wedge. In addition, we find several maxima along the edge that shift towards the apex with increasing loss energy. 
These observations can be understood in the following way\cite{Yalunin2016}: The incident electrons excite SPP wave packets propagating along the edge away from the impact position in positive and negative $z$ direction. If the distance from the impact position to the nano-wedges's apex is not too large, i.e., small compared to the SPP decay length, reflection of the SPPs at the apex leads to  the formation of a standing-wave interference pattern. Hence, we observe for each electron loss energy, a set of EELP maxima along the edge of the nano-wedge.
With increasing electron loss energy the period of this standing wave pattern decreases. Moreover, propagation losses of the SPP result in a reduction of the modulation of the standing wave pattern with increasing distance to the apex.

\begin{figure}[htb]
\centering\includegraphics[scale = 0.83333]{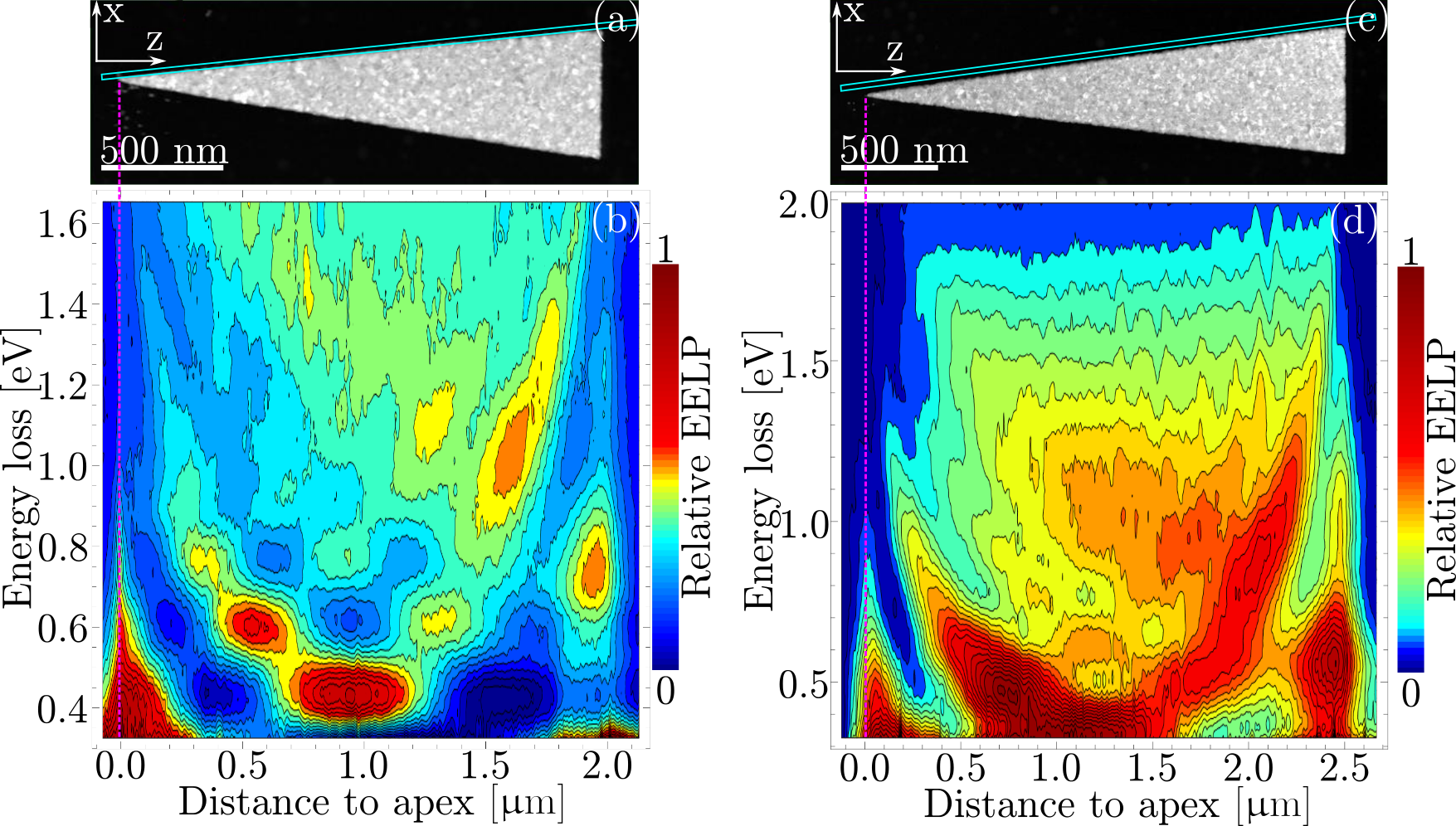}\caption{(a) HAADF micrograph of a $\unit[2.0]{\upmu m}$ long nano-wedge. (b) Relative EELP recorded along the edge of the $\unit[2.0]{\upmu m}$ long nano-wedge (see blue box in the micrograph). (c) HAADF micrograph of a $\unit[2.5]{\upmu m}$ long nano-wedge. (d) Relative EELP recorded along the edge of the $\unit[2.5]{\upmu m}$ long nano-wedge (see blue box in the micrograph).}
\label{225}
\end{figure}

Naturally the question arises for which nano-wedge size a transition from a set of discrete LSPs to a continuum of SPPs takes place in the considered energy range.
By investigating nano-wedges with different sizes, we find that the crossover between these two regimes occurs for a length between $\unit[2.0]{\upmu m}$ and $\unit[2.5]{\upmu m}$.
Figure~\ref{225}(b) depicts the relative EELP recorded parallel to the edge of a $\unit[2.0]{\upmu m}$ long nano-wedge. 
This nano-wedge clearly shows a resonant behavior with several distinct LSP modes. In contrast, a qualitatively different behavior is observed for a $\unit[2.5]{\upmu m}$ long nano-wedge.
In addition to the maxima at the ends, the corresponding EELP (see Figure~\ref{long}(d)) features two pronounced continuous bands.
With increasing electron energy loss, they shift towards the apex and base of the wegde, respectively. These bands can be interpreted as maxima of the two standing wave patterns resulting from the reflection of SPPs at the apex and the base, respectively.

\begin{figure}[hbt]
\centering\includegraphics[scale = 0.83333]{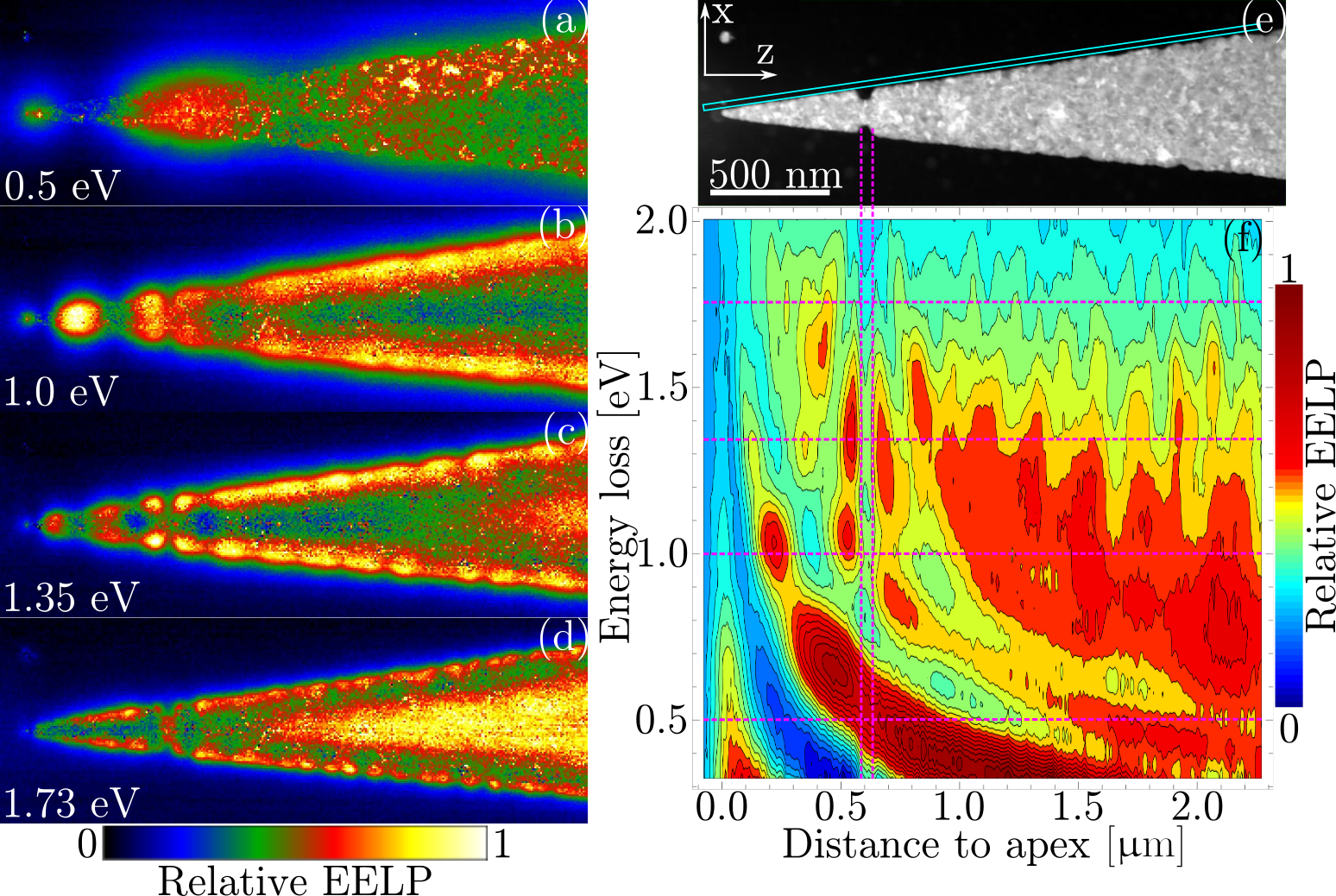}\caption{
(a) -- (d): EEL maps of a $\unit[15]{\upmu m}$ long gold nano-wedge with grooves for electron loss energies of $\unit[0.5]{eV}$, $\unit[1.0]{eV}$, $\unit[1.35]{eV}, $and $\unit[1.73]{eV}$. (e): HAADF micrograph of the investigated area of the structure. The blue box indicates the area where the EELP data in (f) has been recorded. Red doted lines in (f) indicate the position of the grooves and the energies of the shown EEL maps.
}\label{grooves}
\end{figure}

Adding grooves to a long nano-wedge leads to interesting new effects. Figure~\ref{grooves}(e) shows the apex region of a $\unit[15]{\upmu m}$ long nano-wedge. $\unit[600]{nm}$ away from the nano-wedge's apex two $\unit[50]{nm}$ broad and $\unit[55]{nm}$ deep grooves are cut, leaving a fillet with a width of $\unit[100]{nm}$, separating the front section from the rear section. The front section has the same dimensions as the short wedge discussed above. 
For low electron loss energies, one expects that the grooves have a minor effect on the SPP propagation. Hence, the structure should behave similar to the long nano-wedge without grooves. 
Our experimental data confirms this prediction. The EEL map with an electron loss energy of $\unit[0.5]{eV}$ (see Figure~\ref{grooves}(a)), as well as the low energy region ($<\unit[0.85]{eV}$) of the EELP distribution recorded along the edge (see Figure~\ref{grooves}(f)) show a clear SPP like behavior. In particular, we observe that the maxima in the EELP distribution continuously shift towards the apex with increasing energy loss. This is a clear signature of the formation of a standing wave pattern as also observed in figure~\ref{long}. 
In contrast to this, for loss energies above approximately $\unit[0.85]{eV}$, a clear influence of the grooves can be observed. In this case, the front section of the structure ($z<\unit[600]{nm}$) shows a similar behavior as the small nano-wedge. More specifically, we can identify two distinct resonances in the front section at $\unit[1.0]{eV}$ and $\unit[1.35]{eV}$ with three and four maxima, respectively. These resonances correspond to the second and third LSP mode of the short nano-wedge (cf. figure~\ref{short}).
Furthermore, we observe in the EELP distribution a pronounced maximum at the right side of the grooves for loss energies above $\unit[0.85]{eV}$. Additionally, a second maximum occurs in the rear section that continuously shifts towards the grooves with increasing electron loss energy. This indicates the formation of a standing wave pattern, where the reflection of the SPPs takes place at the grooves.

\begin{figure}[htb]
\centering\includegraphics[scale =0.83333]{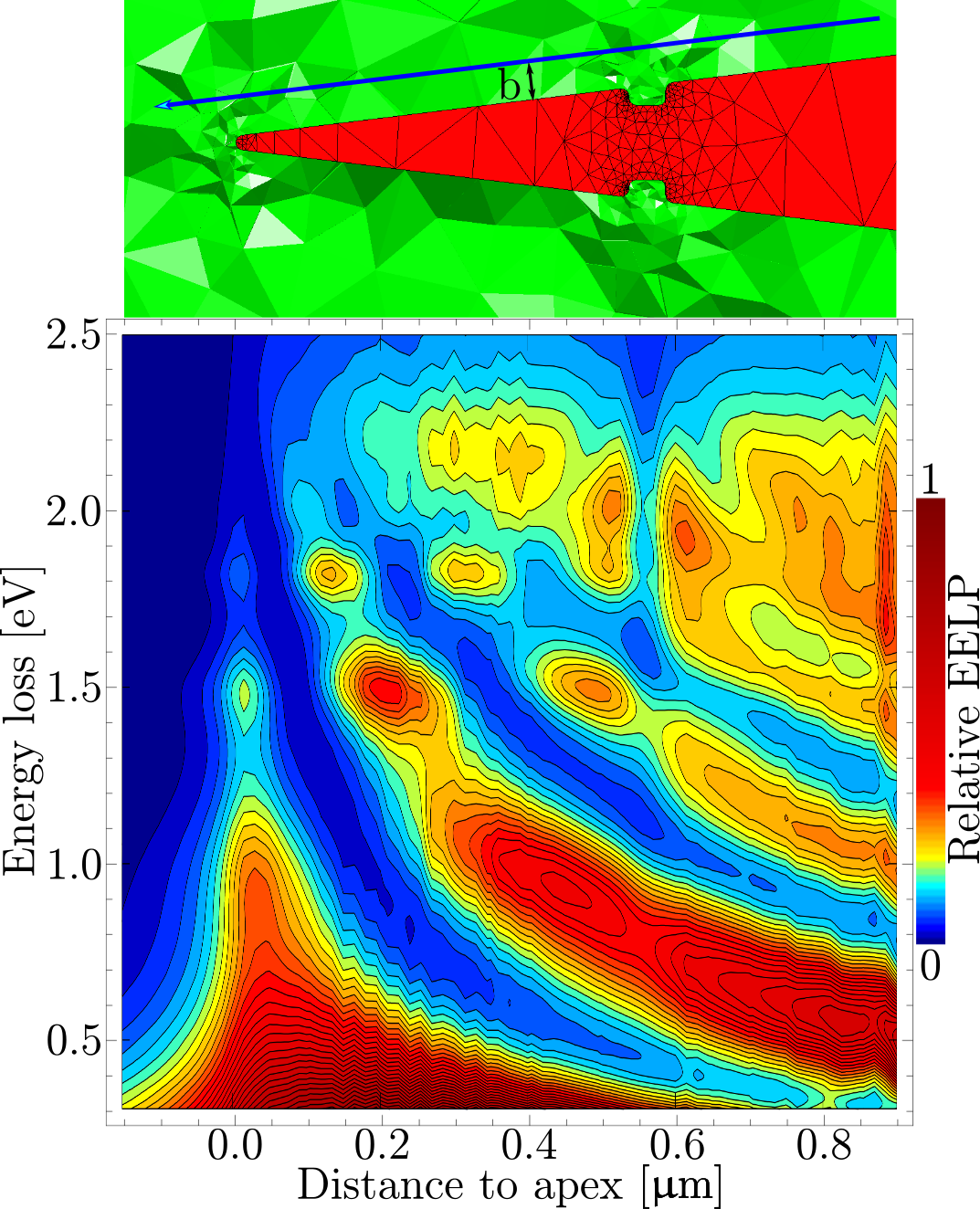}\caption{Upper part: Mesh of the grooved nano-wedge used for DGTD method computations. Lower part: computed relative EELP color coded in dependence of distance to apex and energy loss.}
\label{fig:dg_mesh_and_results}
\end{figure}

Figure~\ref{fig:dg_mesh_and_results} depicts a cross-section of the three-dimensional mesh of the modified nano-wedge used in the DGTD computations.
The EELP is computed along a line parallel to one edge of the nano-wedge (see blue line in upper part of figure~\ref{fig:dg_mesh_and_results}). The impact parameter $b$ between line and edge of the nano-wedge is $\unit[10]{nm}$ and the line does not follow the shape of the groove. The computed EELP data (see figure~\ref{fig:dg_mesh_and_results}, lower part)  qualitatively reproduces the experimental results.
At low electron loss energies a clear signature of SPPs running towards the apex is observed. For electron loss energies exceeding approximately $\unit[1.2]{eV}$, distinct maxima in the EELP indicative of LSPs occur at $\unit[1.5]{eV}$, $\unit[1.8]{eV}$, and $\unit[2.1]{eV}$ in the front section of the structure. 
As before, the deviation in energy of the computations from the experimental data can be most likely traced back to deviations of the assumed geometry from the real sample parameters and an energy shift caused by the underlying substrate. 

\section{Conclusion}
In this work we demonstrated the transition from  LSP-supporting nano-structures to a SPP-supporting structures by prolongating the structural dimensions from some hundred nanometers to some ten micrometers. For the shorter samples, multiple reflections at the ends lead to the formation of LSPs. For the longer samples, also propagating plasmons can be found. These are SPPs that are reflected at one end of the structure. The complex behavior of a modified sample shows that the addition of sub-wavelength features to extended nano-structures can result in a rich interplay of localized LSPs and delocalized SPPs within the same energy range.

\section*{Funding}
S.L, S.I., and T.W. gratefully acknowledge financial support by the Deutsche Forschungsgemeinschaft project LI 1641/5-1.
K.B. and T.K. acknowledge support by the Einstein Foundation within the project "ActiPLAnt".
\end{document}